\documentclass[prl,floats,twocolumn]{revtex4-1}
\usepackage{graphicx}
\usepackage{color}
\usepackage{amssymb, amsmath, amsbsy}

\begin{document}

\title{Macroscopic degeneracy of zero-mode rotating surface states in 3D Dirac 
   and Weyl semimetals under radiation}

\author{Jos\'e Gonz\'alez}
\author{Rafael A. Molina}

\affiliation{Instituto de Estructura de la Materia, IEM-CSIC, Serrano 123, Madrid 28006, Spain}



\begin{abstract}
We investigate the development of novel surface states when 
3D Dirac or Weyl semimetals are placed under circularly polarized
electromagnetic radiation. 
We find that the hybridization between inverted Floquet bands opens 
in general a gap, which closes at so-called exceptional points found 
for complex values of the momentum. This corresponds to the 
appearance of midgap surface 
states in the form of evanescent waves decaying from the surface 
exposed to the radiation. We observe a
phenomenon reminiscent of Landau quantization by which the midgap 
surface states get a large degeneracy proportional to the radiation 
flux traversing the surface of the semimetal.
We show that all these surface states carry angular current, 
leading to an angular modulation of their charge that rotates with the 
same frequency of the radiation, which should manifest in the observation 
of a macroscopic chiral current in the irradiated surface.
\end{abstract}



\maketitle
{\em Introduction.---}
In recent years we have witnessed the discovery of several types of 
materials characterized by having electron quasiparticles with linear 
momentum dispersion.
Graphene was certainly the first of those 
materials\cite{novo}, but afterwards we learned about the topological 
insulators\cite{top1,top2}, 
to end up more recently with the investigation of 3D semimetals whose 
low-energy excitations behave as Dirac\cite{liu,neupane,borisenko} or Weyl 
fermions\cite{taas1,taas2}.

These materials have attracted a lot of attention for their potential
to develop a new type of electronic transport without dissipation. The key 
idea is that of topological protection, which 
has its precedent in the edge states of the quantum Hall effect. 
The surface states in the novel materials may 
also have a well-defined chirality, protecting them against 
backscattering. 
Both properties are fundamental for revolutionary applications in spintronics and fault-tolerant quantum computation \cite{top1,top2}. 

In this search, the interaction between light and matter can play an important
role, since the electromagnetic radiation may be a versatile resource to change and 
control topological states of matter. In 2D semimetals, it can open  
a gap in the bulk, leading to chiral currents at the boundary of the 
electron system\cite{oka,refael,inoue,demler,fu,arovas,calvo11,gomez13,perez,calvo15},  
as observed on the surface of 3D topological insulators\cite{jar}. 
The effect of the radiation has been also investigated in the case of 3D Dirac 
and Weyl semimetals, focusing on bulk properties\cite{Wang14,nara,palee}.
    
\begin{figure}[h]
\begin{center}
\includegraphics[height=4.2cm]{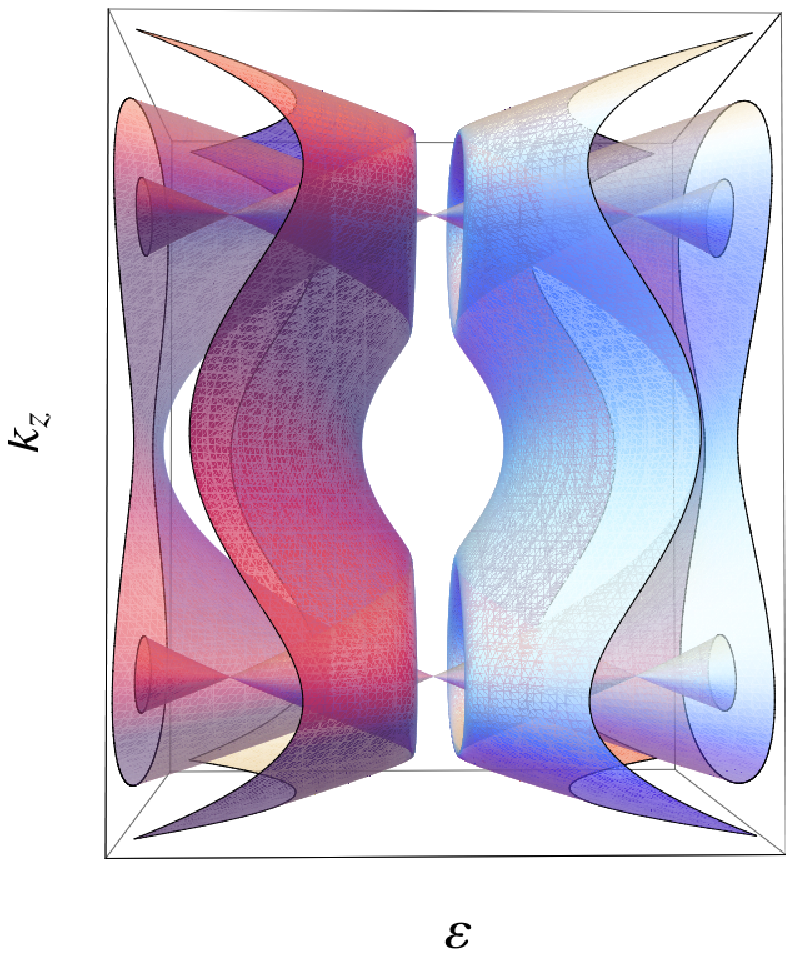}
\hspace{1.5cm}
\includegraphics[height=4.5cm]{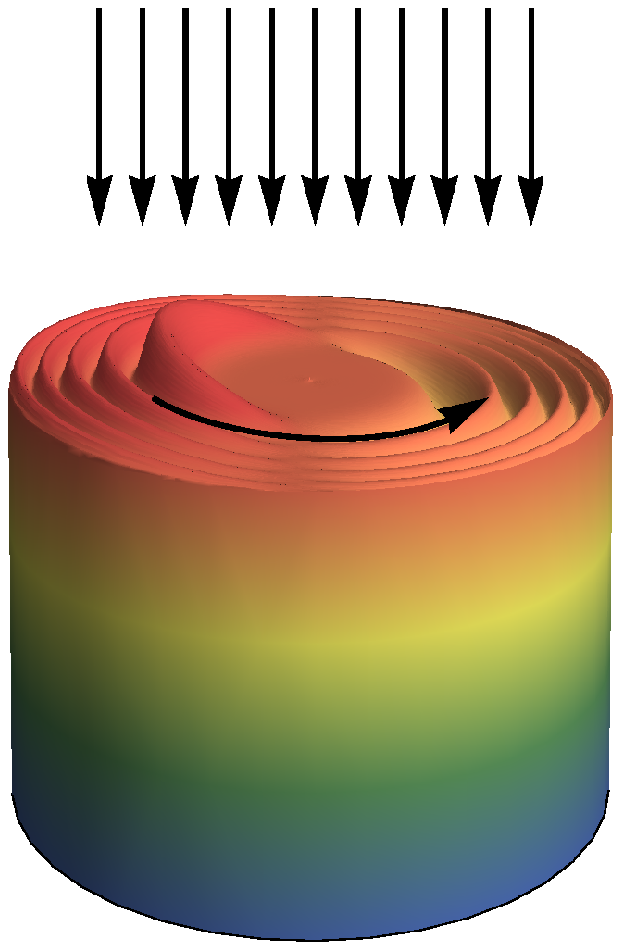}
\end{center}
\caption{Left: Schematic representation of the first Floquet bands with quasi-energy
$\epsilon $ and $\epsilon \pm \hbar \Omega $ ($\Omega $ being the 
radiation frequency) for a model of 3D semimetal with two nodes along the 
momentum axis $k_z$ (red color denotes positive energy while light blue denotes negative 
energy, the direction transverse to $k_z$ represents additional components of the momentum). 
Right: Geometry of 3D semimetal with the surface exposed 
to the radiation where the evanescent states appear.}
\label{one}
\end{figure}

In this paper, we investigate the development of novel surface states
when a 3D Dirac or Weyl semimetal is placed under circularly polarized 
electromagnetic radiation. 
We will show that such states are intimately related to avoided crossings at 
the gap that opens up from the hybridization of inverted Floquet bands, 
as represented in Fig. \ref{one}. 
The gap closes at so-called exceptional points (EPs) \cite{berry,heiss},
which are celebrated in the context of non-Hermitian Hamiltonians and 
appear here for complex values of the momentum describing the evanescence 
into the 3D semimetal. 
The stability of the novel surface states is then guaranteed by a new 
mechanism of topological protection, relying on the fact that each EP
comes as a branch point in the spectrum which cannot be removed 
by small perturbations. 


The genuine feature of the novel surface
states is that they all carry a significant angular current, with the same chirality of the photon polarization and the frequency of the radiation. 
Such states may prove especially useful in the current drive towards increasing the frequency limits of electronic devices \cite{Krausz14}.

{\em Model and Floquet theory.---}
A low-energy Hamiltonian for a 3D Dirac semimetal around the Brillouin zone center, considering terms
up to quadratic order in the quasimomentum $\mathbf{k}$, can be written as 
\cite{Wang12,Wang13}.
\begin{eqnarray}
H & = & \epsilon_0(\mathbf{k})\mathbb{I}+M(\mathbf{k})\sigma_z + \hbar v(\zeta k_x\sigma_x-k_y\sigma_y) \label{eq;H0} \\
\epsilon_0(\mathbf{k}) & = & c_0+c_1k_z^2+c_2(k_x^2+k_y^2) \\
M(\mathbf{k}) & = & m_0-m_1 k_z^2-m_2(k_x^2+k_y^2),
\end{eqnarray}
where $\mathbb{I}$ is the $2 \times 2$ identity matrix, $\sigma_i, \, i=x,y,z$, are the Pauli matrices, and $\zeta= \pm 1$ sets the chirality in the Dirac cones.
The solution of the eigenvalue problem is
\begin{equation}
E_{\pm}=\epsilon_0(\mathbf{k})\pm \sqrt{M(\mathbf{k})^2 + \hbar^2 v^2(k_x^2+k_y^2)}.
 \label{eq:bands}
\end{equation}
With the parameters $m_0,m_1,m_2 < 0$ to reproduce band inversion, the spectrum shows Dirac crossings at $\mathbf{k}_c=(0,0,\pm\sqrt{m_0/m_1})$. Ignoring the part proportional to the unit matrix $\epsilon_0(\mathbf{k})$, we can expand the Hamiltonian linearly around each Dirac point to
obtain a model for 3D massless Dirac fermions with anisotropic linear dispersion
$E(\mathbf{k})= \pm \sqrt{\hbar^2 v^2(k_x^2+k_y^2) + 4 m_0 m_1 (k_z - k_{c,z})^2}$. 

Consider illumination by circularly polarized off-resonant light of frequency $\Omega$ and field amplitude $\mathcal{F}$.
In the case of polarization in the $x-y$ plane, light produces the vector potential in the dipolar approximation
 $\mathbf{\mathcal{A}}(t)=\mathcal{A}(\eta \sin{\Omega t},\cos{\Omega t},0)$,
where $\eta=\pm 1$ for right and left circularly polarized beams, respectively, and $\mathcal{A}=\mathcal{F}/\Omega$. 
We make the Peierls substitution $\mathbf{k} \rightarrow \mathbf{k}+\mathbf{\mathcal{A}}(t)$
and use Floquet theory in order to compute the band structure in the presence of the radiation field\cite{supp}. Solutions of the time-dependent Schr\"odinger equation in the case of time-periodic Hamiltonians have the form $\left|\Psi(t)\right>=e^{-i\epsilon t/\hbar} \left|\Phi(t)\right>$, with a conserved quantity, the quasi-energy $\epsilon$, playing a similar role to the energy in the time-independent Sch\"odinger equation. The Floquet states $\left|\Phi(t)\right>$ are periodic in time with the same period as the Hamiltonian and they can be developed in Fourier series, 
$\left|\Phi(t)\right>=\sum_m e^{-im\Omega t}\left|u_{\alpha}^m\right>$. 
This transforms the time-dependent Schr\"odinger equation into 
\begin{equation}
 \sum_n H^{mn} \left|u_{\alpha}^n\right>=\left(\epsilon_{\alpha}+ m \hbar \Omega\right)\left|u_{\alpha}^m\right>,
 \label{eq:Floquet}
\end{equation}
where the Floquet Hamiltonian matrix elements are given by
 $H^{mn}=\frac{1}{T}\int^T_0 dt H(t) e^{i(m-n)\Omega t}$,
$T=2\pi/\Omega$ being the period of the time-dependent Hamiltonian\cite{supp}.

\begin{figure}
\includegraphics[width=0.23\textwidth]{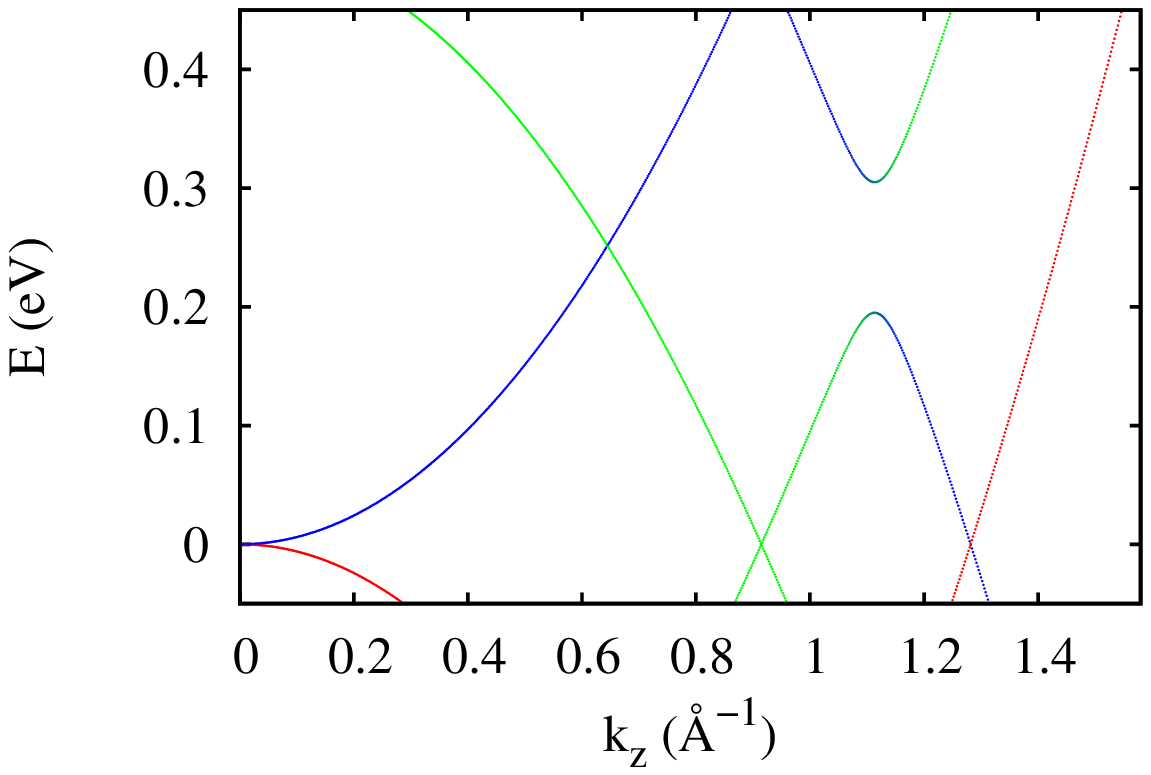}
\includegraphics[width=0.23\textwidth]{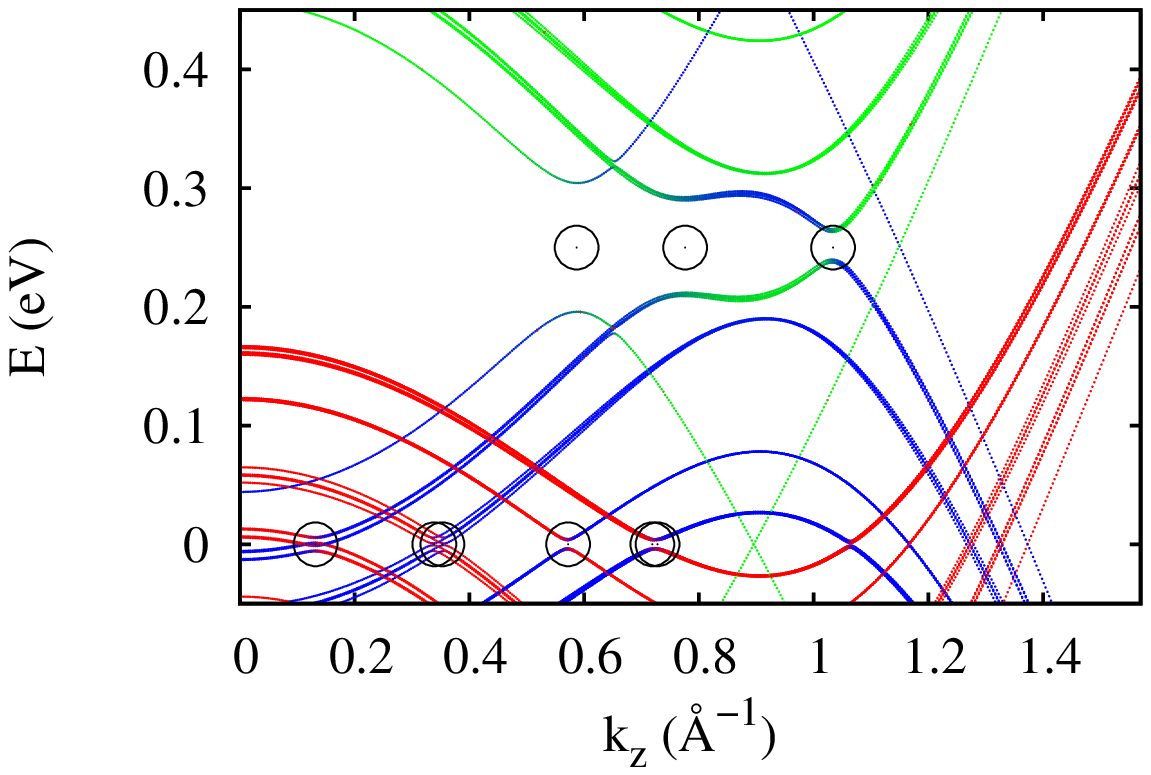}
\caption{\label{fig:xywire} Band structure for 
model parameters $c_0=c_1=c_2=0.0$, $m_0=-0.5$ eV, $m_1=-0.605$ eV \AA$^2$, $m_2=-1.0$ eV \AA$^2$, $\hbar v=1.1$ eV \AA, $\hbar \Omega=0.5$ eV, $\mathcal{A}=0.05$ \AA$^{-1}$, in the bulk (left panel) and for a wire of section $S=150 \times 150$ \AA$^2$ in
the $x-y$ plane (right panel). The position of the Dirac cone at zero field is
$k_{c,z}=\sqrt{m_0/m_1}=0.909$ \AA$^{-1}$. The color represents the predominant Floquet mode of the particular state in a RGB coding where the intensity of each mode is mapped to the intensity of each fundamental color, $n=-1,0,1$ correspond to red, green and blue, respectively.  The position of the evanescent waves obtained as strictly real solutions of the eigenvalue problem after including an imaginary part in $k_z$ of Im$(k_z)=0.01$ \AA$^{-1}$ is shown with black circles with a dot.
A few high-energy bands with $n=-1$ index not affecting the avoided crossing structures have been erased for clarity of presentation.}
\end{figure}

As a relevant example, we perform calculations in an infinite wire in the $z$ direction with a finite square section in the $x-y$ plane.
Using the Floquet formalism, we calculate the band structure as a function of $k_z$ discretizing the model Hamiltonian with a standard tight-binding regularization including the Floquet subbands.
Results with parameters of the model in the topological Dirac semimetal regime are shown in
Fig. \ref{fig:xywire}, where we compare the energy of the bulk bands (left) with the quasi-energy of the bands in the wire (right) as a function of $k_z$. The structure of the bands wrapping up in the Floquet-Brillouin zone is very apparent. In the middle of the Floquet-Brillouin zone at $\epsilon=0$, a gap opens between states with Floquet modes $n=-1$ (red) and $n=1$ (blue), the hybridization being a second order effect leading to a quite small gap, proportional to $\mathcal{A}^2$. The much bigger gap (proportional to $\mathcal{A}$) between modes $n=0$ (green) and $n=1$ is in one of the edges of the Floquet-Brillouin zone at $\epsilon=\hbar \Omega/2$, with a mirror structure (not shown) at $\epsilon=-\hbar \Omega/2$ between modes $n=0$ and $n=-1$. 

In order to explore the possibility of evanescent states in the $z$ direction, we solve the Schr\"odinger equation in the same geometry but including a finite imaginary part in $k_z$. The Hamiltonian becomes then non-Hermitian and the eigenenergies acquire a finite imaginary part. These are obviously non-physical solutions of the problem. However, non-Hermitian Hamiltonians may have eigenenergies that are strictly real, in the form of EPs \cite{berry,heiss}. These correspond to physical evanescent solutions with the imaginary part of the momentum giving the decay length of a wave in a semi-infinite geometry. 
The black circles in Fig. \ref{fig:xywire} mark the presence of the evanescent states in the gaps induced by the radiation. Not all the avoided crossings present evanescent states, that depends on the properties of the actual bands involved. However, the number of avoided crossings is proportional to $\mathcal{A}$ and to the area of the section of the wire.
This may imply in general a large degeneracy of evanescent states, whose origin is clarified in the continuum limit discussed in what follows. 


{\em Quantization and localization of zero-mode surface states.---}
In order to unveil the properties of the novel surface states, we focus now 
on the low-energy physics around any of the nodes of the Hamiltonian 
(\ref{eq;H0}), taking moreover the projection for a given chirality $\zeta$.
This amounts to linearize $M(\mathbf{k})$ about the node at 
$\mathbf{k} = \mathbf{k}_c$, which leads in real space to the Hamiltonian for a 
single Weyl quasiparticle 
\begin{eqnarray}
H_W  & = &  -i \hbar (v \sigma_x \partial_x + v \sigma_y \partial_y  
                    + v_z \sigma_z \partial_z )        \nonumber        \\
 &  & + \hbar v \mathcal{A} \left(\sigma_x  \cos (\Omega t) + \sigma_y  \sin (\Omega t) \right)
\label{ham}
\end{eqnarray}
with a velocity $v_z \neq v$ along the $z$-axis\cite{vz}.

$H_W$ can be translated into a time-independent Hamiltonian after applying the 
unitary transformation $U =  e^{-i J_z \Omega t/\hbar }$, with the 
projection of the total angular momentum 
$J_z = -i\hbar x \partial_y + i\hbar y \partial_x + \hbar \sigma_z/2$ 
\cite{lopez}. This leads to the transformed Hamiltonian
\begin{eqnarray}
\widetilde{H}_W   & = & U^\dagger H_W U - i \hbar U^\dagger \partial_t U   \nonumber      \\
  & = &  -i \hbar ( v \sigma_x \partial_x + v \sigma_y \partial_y  
                    + v_z \sigma_z \partial_z )
      -  \Omega J_z  + \hbar v \mathcal{A} \sigma_x      \;\;\;\;
\label{ham2}
\end{eqnarray}
Each eigenvector $\chi $ of (\ref{ham2}) corresponds then to 
a solution $\Psi (t)$ of the original time-dependent Schr\"odinger equation, 
given by 
$\Psi (t) = e^{-i J_z \Omega t/\hbar } e^{-i\varepsilon t/\hbar} \chi $ 
($\varepsilon $ being the eigenvalue of (\ref{ham2})). This shows that the 
eigenvalues $j_z$ of the projection $J_z$ can be used in this approach to 
label the different side bands arising from the irradiation.   
 
After applying the 
gauge transformation $P = \exp (-i \mathcal{A} x )$ to (\ref{ham2}), the 
Hamiltonian becomes (in polar coordinates $r, \theta$ and with 
$\sigma_{\pm } = (\sigma_x \pm i \sigma_y)/2$)
\begin{eqnarray}
\widetilde{H}'_W & = &  -i \hbar v \sum_{s = \pm } e^{-s i \theta }
 \left(\partial_r - s i \frac{1}{r} \partial_\theta \right) \sigma_s   \nonumber  \\
   & &  -i \hbar v_z \sigma_z \partial_z  
        -  \hbar \Omega \left( -i \partial_\theta + \frac{\sigma_z}{2} \right)   
         - \hbar \Omega \mathcal{A} r \sin(\theta)    \;\;\;\;
\label{ham3}
\end{eqnarray}
The last term in (\ref{ham3}) is responsible for the coupling between states 
with different angular momenta $j_z$. Then, the avoided 
crossing of the bands with $j_z = \pm \hbar/2$ gives rise to a gap around zero 
energy in the spectrum of (\ref{ham3}). This point corresponds actually to a 
quasi-energy $\epsilon = \hbar \Omega /2$ in the conventional Floquet approach, 
since $U$ has the effect of shifting the energy of the side bands in the 
present approach by half-integer multiples of $\hbar \Omega $. 

\begin{figure}[h]
\begin{center}
\includegraphics[height=3.5cm]{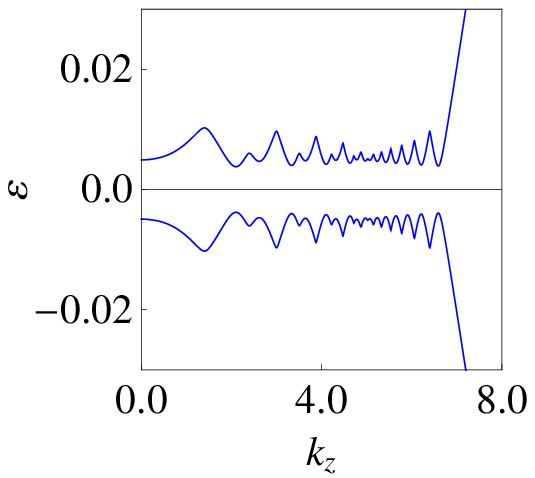}
\includegraphics[height=3.65cm]{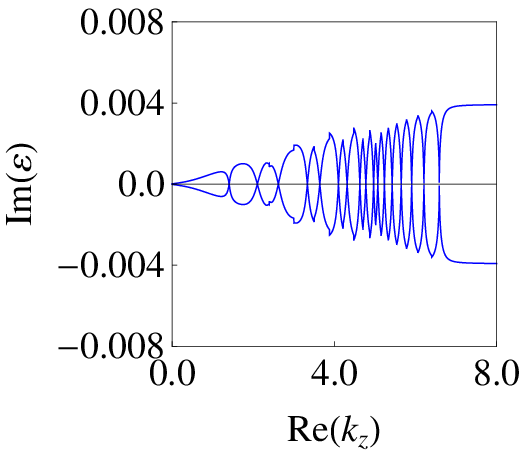}
 \mbox{} \hspace{8.0cm}  (a) \hspace{3.5cm} (b) 
\end{center}
\caption{(a) Plot of the gap of the Hamiltonian (\ref{ham3}) (in eV) 
as a function of real $k_z$ (measured in units of 
the inverse of the typical microscopic length scale $a$ in the material) for 
$\hbar v/a = 1.1$ eV, $\hbar \Omega = 0.5$ eV, $\mathcal{A} = 0.005 \: a^{-1}$, 
and radius $R = 200 a$ of the cylindrical geometry considered in the text.
(b) Imaginary part of the lowest eigenvalue of the 
Hamiltonian (\ref{ham3}) (in eV) as a function of ${\rm Re} (k_z)$, for complex
momenta with ${\rm Im} (k_z) = 0.08 \: a^{-1}$ and the same
parameters as in (a).}
\label{four}
\end{figure}

The development of the gap can be captured (for not too large amplitude 
$\mathcal{A}$) by solving in the space spanned by 
a linear combination of spinors with projection of the total angular momentum 
$j_z = \pm \hbar/2$,
\begin{equation}
\chi = \left( \begin{array}{cc} \phi_1 (r) 
              \\ e^{i\theta} \phi_2 (r) \end{array} \right) e^{i k_z z} + 
  \left( \begin{array}{cc} e^{-i\theta} \phi_3 (r) 
             \\  \phi_4 (r) \end{array} \right)  e^{i k_z z}
\label{spin}
\end{equation}
We have considered specifically a cylindrical geometry with the section at 
$z = 0$ exposed to the radiation, as represented in Fig. \ref{one} \cite{f2}. 
A typical shape of the gap is shown in Fig. \ref{four}(a), 
where we see that it has several oscillations before recovering the linear 
dispersion of the original cones beyond a certain $k_z$. 
This is indeed the generic behavior, with an increasing number of 
oscillations as the radius $R $ of the cylinder grows.

The origin of the oscillations in the gap is clarified by noting that they
arise from the existence of EPs in the spectrum of the Hamiltonian 
(\ref{ham3}), corresponding to values of $k_z$ inside the complex 
plane where the gap closes. Around each oscillation of the gap, the lowest 
eigenvalue $\varepsilon $ behaves as a function of complex $k_z$ as
$\varepsilon \sim \sqrt{k_i - k_z} \sqrt{\overline{k}_i - k_z}$, with EPs at 
complex conjugate momenta $k_i, \overline{k}_i$ \cite{supp}. Evanescent states 
with ${\rm Im} (\varepsilon ) = 0$ exist then in the segment 
$|{\rm Im} (k_z)| \leq |{\rm Im} (k_i)|$, for each pair $k_i, \overline{k}_i$. 
We have represented for 
instance in Fig. \ref{four}(b) the behavior of the imaginary part of the 
lowest eigenvalue, when ${\rm Im} (k_z) \neq 0$, as a function of 
the real part of $k_z$. We 
observe the recurrent development of complex momenta at which the eigenvalue 
becomes purely real, leading to a set of evanescent eigenstates in perfect 
correspondence with the minima of the gap in Fig. \ref{four}(a).

We find then that the evanescent states are preserved by a mechanism of
topological protection, as the branch cuts cannot be undone in the complex 
plane unless the branch points coalesce in pairs. It can be seen that the 
imaginary part of the EPs has a very smooth dependence on the frequency 
$\Omega $, while it grows linearly with the amplitude $\mathcal{A}$ \cite{f1}. 
The number of zero-mode evanescent states may be actually characterized from 
the number of branch points that the lowest band $\varepsilon (k_z)$ has in 
the complex plane.

\begin{figure}[h]
\begin{center}

\begin{tabular}{cc}

\includegraphics[height=4.2cm]{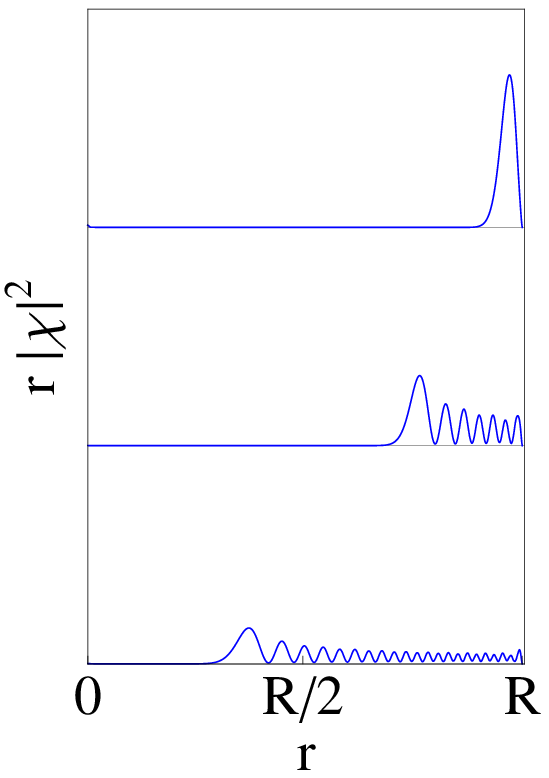}  &
\hspace{0.5cm}
\raisebox{2.0cm}{
\begin{tabular}{c}
\includegraphics[height=1.9cm]{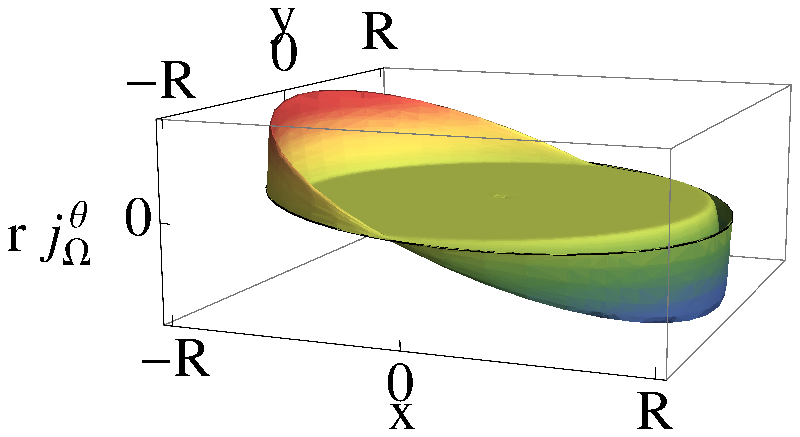}  \\
\includegraphics[height=2.0cm]{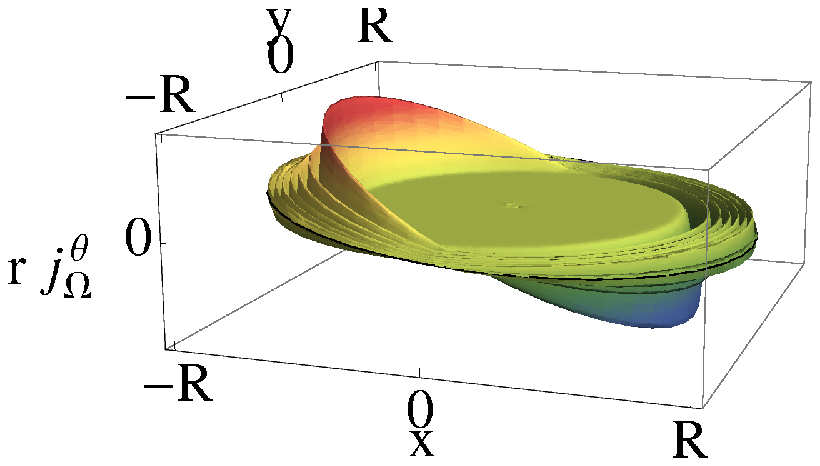}  
\end{tabular}    }

\end{tabular}

\end{center}
\caption{Left: Probability distribution $r |\chi |^2$
along the radial direction for evanescent states corresponding to three 
decreasing values of ${\rm Re} (k_z)$ with ${\rm Im}(\varepsilon ) = 0 $ 
(starting with the outermost evanescent state) for the same parameters as in 
Fig. \ref{four} and $R = 400a$. 
Right: Time-dependent angular current $r j_\Omega^\theta $ for the states in 
the upper part (top) and the middle part (bottom) of the left panel, 
represented for fixed time $t = 0$ at the irradiated surface of the cylindrical 
geometry.}
\label{six}
\end{figure}

It can be checked that the number of EPs increases as the radius $R$ grows,
leading to a definite pattern of quantization in the surface of our geometry. 
It can be seen that the order of each zero in the plot of Fig. \ref{four}(b) 
(from right to left) gives also the extent of the localization that the 
corresponding evanescent wave has in the radial direction, as illustrated in 
Fig. \ref{six}. The peak in the probability distribution shifts to lower 
values of $r$ as ${\rm Re} (k_z)$ decreases, reaching eventually the innermost 
region of the surface. The number of evanescent states, besides growing 
linearly with the frequency $\Omega $ and the amplitude $\mathcal{A}$, becomes
actually proportional to the area of the irradiated surface, at a rate of 
roughly one state per $(100a)^2$ (for $\hbar \Omega = 0.1$ eV, 
$\mathcal{A} = 0.01 \: a^{-1}$, $a$ being a typical microscopic length scale 
in the material).

Furthermore,
the midgap evanescent states evolve in time with a rotation of their 
charge along the angular variable $\theta$. This can be shown by computing the 
angular component $j^\theta$ of the probability current for the Hamiltonian 
(\ref{ham}). This current has a static contribution, 
$j_{\rm static}^\theta = -i (v/r) (\phi_1^* \phi_2 + \phi_3^* \phi_4 ) 
 + {\rm h.c.}$, which leads to a very small intensity when integrated over the
radial direction, as a consequence of the tendency of the two 
contributions from $j_z = \pm \hbar/2$ states to cancel each other\cite{supp}. 
For $\hbar \Omega = 0.5$ eV and $\mathcal{A} = 0.005 \: a^{-1}$, we get for
instance values of the intensity $\int dr \: r j_{\rm static}^\theta $ between 
$\sim 10^{-5} v/a$ and $\sim 10^{-4} v/a$, when computing for evanescent states 
from the outermost to the innermost region of the surface. However, there is 
also a non-negligible time-dependent contribution to $j^\theta$ \cite{supp}, 
given by
\begin{equation}
j_\Omega^\theta = -i (v/r) (\phi_1^* \phi_4 e^{-i(\theta - \Omega t)} 
                + \phi_3^* \phi_2  e^{i(\theta - \Omega t) } ) + {\rm h.c.}
\label{jomega}
\end{equation}
which has a periodic dependence on $\theta - \Omega t$ as shown in Fig. 
\ref{six} \cite{f3}. The intensity corresponding to $j_\Omega^\theta $ has 
maxima (in the angular variable) which turn out to be in general about two 
orders of magnitude above the intensity obtained from $j_{\rm static}^\theta $, 
for every evanescent state\cite{supp}.

When introduced in the continuity equation, the form of $j_\Omega^\theta $
implies that the charge of each state must have a periodic modulation in the 
angular variable, and that it must rotate with frequency $\Omega $ along the 
concentric rings where each evanescent state is confined\cite{supp}. From a 
practical point of view, this leads to the formation of rotating dipoles on 
the surface of the system, whose movement can be controlled by tuning the 
parameters of the radiation.

While we have referred here to the hybridization of states with 
$j_z = \pm \hbar/2$, the coupling of side bands with higher values of $J_z$ 
results in surface states with similar properties of localization and time 
evolution\cite{supp}. It can be seen in particular that the hybridization of 
side bands with values of $j_z$ differing by $2\hbar$ leads also to 
evanescent states, which correspond in that case to the EPs found at 
$\epsilon=0$ in the preceding section.


{\em Conclusion.---}
An important practical consideration in this study is
that the infrared radiation may penetrate sufficiently deep into the 3D 
semimetals, given the limited screening in these materials. The penetration 
length $l$ can be estimated as the inverse of the absorption coefficient 
$\alpha $, which is expressed in terms of the dielectric function 
$\epsilon (\Omega )$ as $\alpha = 2\Omega \: {\rm Im} \sqrt{\epsilon (\Omega )}/c$. 
We find for instance that $l \sim 1 \: \mu m$ in the near-infrared 
($\Omega \sim 100 \: {\rm THz}$)\cite{supp}, which is a large enough distance 
to afford the development of the evanescent states.

The magnitude of the component $j_\Omega^\theta $ of the current 
offers good perspectives to observe experimentally the novel surface states.
In our cylindrical geometry, the intensity of the current across the radial 
direction, $I = e \int dr \: r j_\Omega^\theta $, gets maxima (in the angular
variable) that range 
between $\sim 10^{-1} \: \mu A$ and $\sim 1 \: \mu A$, for individual states 
taken from the outermost to the innermost region in the top surface. The total 
intensity could be enhanced by a large additional factor, in a device able to 
measure the contribution of a significant part of the surface states.

A suitable experimental setup (i.e. two electrodes on top of the surface 
of the semimetal) may be able to convert the rotation of the charge in the 
surface states into an electrical current, if the device is made to work as 
a rectenna. This may benefit from recent developments which make possible to 
rectify currents oscillating even at the frequency of visible light\cite{nn}. 
In our case, 
it may greatly help the fact that the rotation of all the surface states is 
synchronized with that of the radiation fields, making easier the observation 
of the macroscopic chiral current that may develop at the irradiated surface.

\acknowledgements

We acknowledge financial support from MICINN (Spain) through grant No. FIS2011-23713 and MINECO (Spain) through grants No. FIS2012-34479 and FIS2014-57432-P.


\newpage

\onecolumngrid

\begin{center}
{\bf SUPPLEMENTAL MATERIAL}
\end{center}

\vspace{0.5cm}

\begin{center}
{\bf Penetration depth of the electromagnetic radiation in 3D Dirac and Weyl
semimetals}
\end{center}

We analyze here the penetration depth of the electromagnetic radiation 
in a 3D semimetal, focusing on absorption due to interband 
transitions between available electronic states. We start
by noting that the refractive index $n$ of the material can be expressed in
terms of the dielectric function $\epsilon (\omega )$ as
\begin{equation}
n = \sqrt{\epsilon (\omega )}
\end{equation}
According to this expression, $n$ may get in general an imaginary part, which
accounts for the absorption of the electromagnetic radiation by the electron
system. 

A measure of the exponential decay of the radiation energy in the material is
given by the absorption coefficient $\alpha $, from which the penetration 
length $l$ can be estimated as $\sim 1/\alpha $. The absorption coefficient can 
be obtained from the refractive index as 
\begin{equation}
\alpha (\omega ) = (2/c) \: \omega \: {\rm Im} \: n(\omega )
\label{alf}
\end{equation}
In the 3D semimetals, the imaginary part of the dielectric function 
$\epsilon (\omega )$ has a smooth dependence on frequency, so that the
leading dependence of $\alpha $ on $\omega $ is 
linear. More specifically, the dielectric function in a 3D Dirac semimetal 
with $N$ Dirac points (or, equivalently, in a 3D Weyl semimetal with $N$ 
pairs of Weyl points) is given at low energies by
\begin{equation}
\epsilon (\omega ) = \frac{N e^2}{6\pi^2 \hbar v} \: 
  \log \left( \frac{\Lambda}{|\omega|} \right)  +  i \frac{N e^2}{12\pi \hbar v}
\label{diel}
\end{equation}
where $\Lambda $ is a high-frequency cutoff (related to the high-energy cutoff
in the linear electronic dispersion) and $v$ is the Fermi velocity (assuming 
for simplicity that the dispersion is isotropic in all directions). 

Taking typical parameters for known 3D semimetals and keeping $N \sim 1$, we 
get from Eq. (\ref{alf}) a penetration length $l \sim 100$ nm for visible 
light, and $l \sim 1 \: \mu$m in the near-infrared (with $\omega = 100$ THz). 
This shows that the use of infrared radiation may become specially convenient, 
for the sake of observing the interaction between electromagnetic radiation 
and the 3D semimetal proposed in the main text.

\vspace{0.5cm}

\begin{center}
{\bf High-frequency effective Hamiltonian for the 3D Dirac semimetal in the presence of
circularly polarized radiation}
\end{center}

Using the following equations
\begin{equation}
 H^{mn}=\frac{1}{T}\int^T_0 dt H(t) e^{i(m-n)\Omega t},
 \label{eq:Floquet_matrix}
\end{equation}
we can compute the Floquet matrix elements of the model for a 3D Dirac semimetal. 
In our model the only non-zero matrix-elements are $H_{nn}$, $H_{+1}=H_{n+1 n}$, $H_{-1}=H_{n-1 n}$.
The matrix elements $H_{+2}=H_{n+2 n}$, $H_{-2}=H_{n-2 n}$ are zero due to cancellations coming from the shape of the circularly polarized pulses.
It is then, easy to see that the diagonal part of the Floquet Hamiltonian is:
\begin{equation}
H_{nn}=H + (n\hbar \Omega+ c_2\mathcal{A}^2)\mathbb{I}-m_2\mathcal{A}^2\sigma_z,
 \label{eq:renH0}
\end{equation}
This is equivalent to a renormalization of the $c_0$ and $m_0$ parameters in the Hamiltonian $H$, which become
$c_0+c_2\mathcal{A}^2$ and $m_0-m_2\mathcal{A}^2$ respectively.
The non-diagonal parts are
\begin{eqnarray}
 H_{+1}&=&\frac{i}{2} \hbar v\mathcal{A}\eta \zeta \sigma_x- \frac{1}{2} \hbar v \mathcal{A} \sigma_y \nonumber \\
 &+&[(c_2-m_2 \sigma_z)i\mathcal{A}\eta k_x +(c_2-m_2 \sigma_z)\mathcal{A} k_y].
 \label{eq:Hp1}
\end{eqnarray}
\begin{eqnarray}
 H_{-1}&=&-\frac{i}{2} \hbar v\mathcal{A}\eta \zeta \sigma_x-\frac{1}{2} \hbar v \mathcal{A}\sigma_y \nonumber \\
 &+&[(m_2 \sigma_z-c_2)i\mathcal{A}\eta k_x +(c_2-m_2 \sigma_z) \mathcal{A} k_y].
 \label{eq:Hm1}
\end{eqnarray}

The effective high-frequency Hamiltonian when only one-photon processes are taken into account can be written as
\begin{equation}
 H_{\mathrm{eff}}=H_{00}+\frac{[H_{-1},H_{+1}]}{\hbar \Omega}.
 \label{eq:highfreqfloquet}
\end{equation}
For the model of a 3D Dirac semimetal used in the paper:
\begin{equation}
H_{\mathrm{eff}}=H_{00}-\eta \frac{\hbar v^2 \mathcal{A}^2}{\Omega} \sigma_z -2 m_2 \eta \frac{v \mathcal{A}^2}{\Omega}(k_x\sigma_x-k_y\sigma_y).
 \label{eq:heff}
\end{equation}
The terms proportional to $\sigma_z$ change the effective mass terms so the new effective $m_0$ is
\begin{equation}
 m_0^{\mathrm{eff}}=m_0-m_2\mathcal{A}^2-\eta\frac{\hbar v^2 \mathcal{A}^2}{\Omega}.
 \label{eq:effm0}
\end{equation}
The terms proportional to $\sigma_{x(y)}$ renormalize the parameter $v$
\begin{equation}
v^{\mathrm{eff}}=v-2m_2\eta \frac{v \mathcal{A}^2}{\hbar \Omega }.
\label{eq:effa}
\end{equation}

\begin{figure}[b]
 \includegraphics[width=0.42\textwidth]{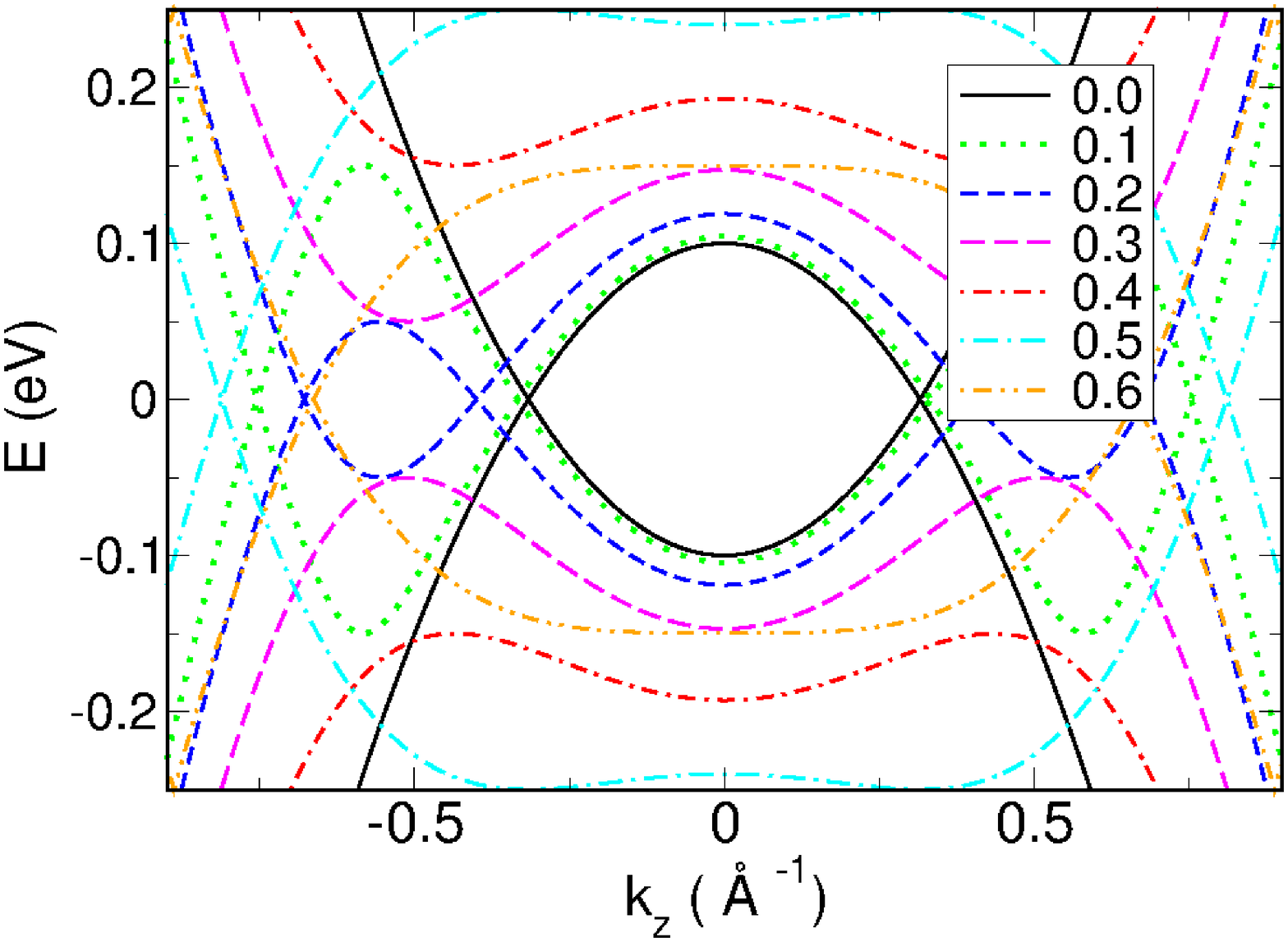}
 \includegraphics[width=0.465\textwidth]{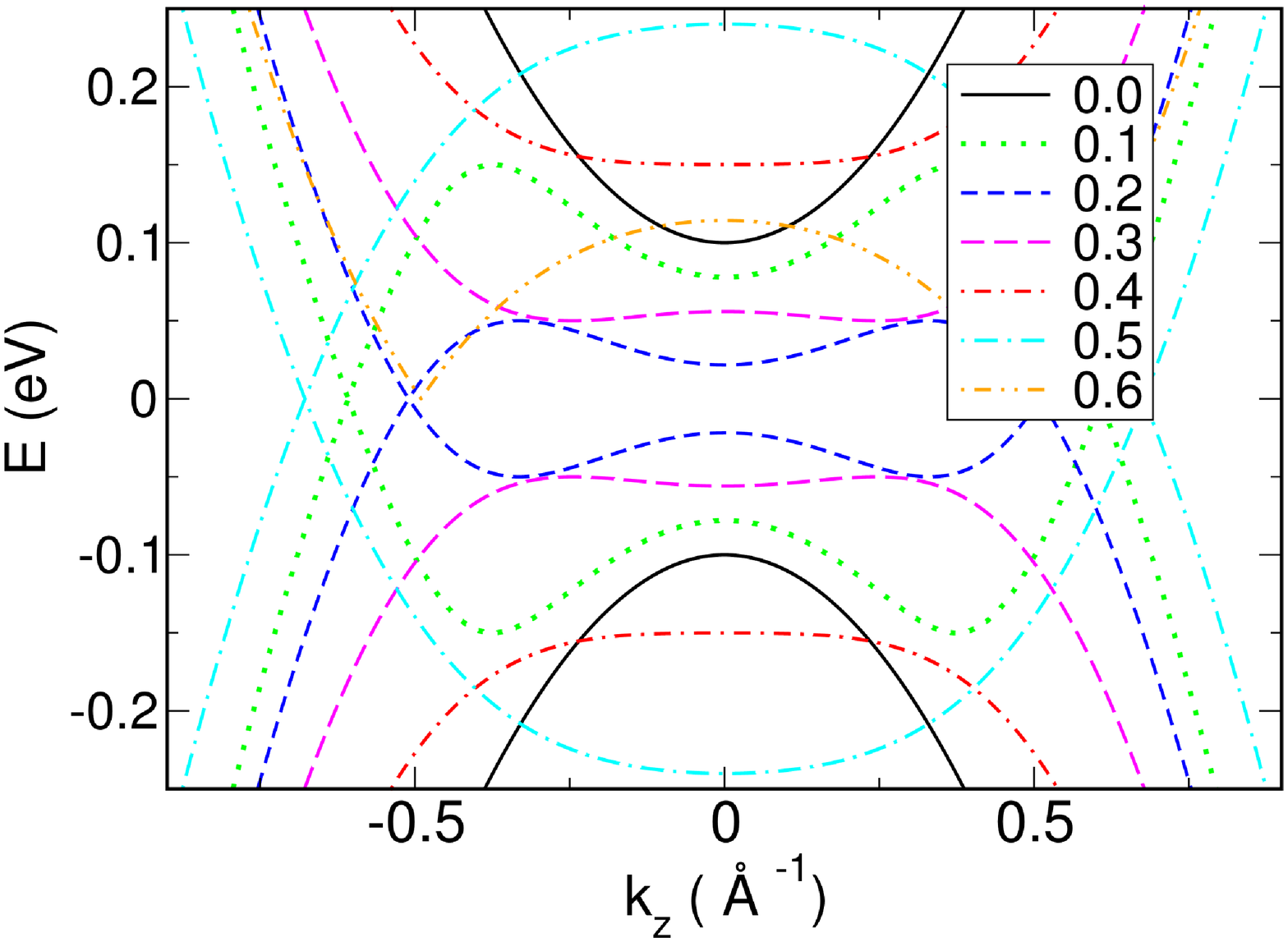}
 \caption{\label{fig:bands} Band structure in the first Floquet-Brillouin zone as a function of $k_z$ with $k_x=k_y=0$. The Hamiltonian parameters are $c_0=c_1=c_2=0.0, m_0=-0.1$ eV (left) $m_0=0.1$ eV (right), $m_1=m_2=1.0$ eV \AA$^2$,  $\hbar v=1.0$ eV \AA, $\zeta=1$, and the external field parameters $\eta=1$, $\hbar \Omega=0.5$ eV, and several values of $\mathcal{A}$ indicated in the legend from $0.0$ to $0.6$ \AA$^{-1}$. In the figure we have eliminated contributions from higher bands at large values of $k_z$ for clarity of presentation.}
\end{figure}

The Dirac crossing points are displaced decreasing or increasing the distance between them depending on the polarization of the external field $\eta$. 
Choosing the parameters of the external field wisely, we can switch the band inversion on or off by changing the sign of $m_0^{\mathrm{eff}}$.
Although the high-frequency approximation fails to properly describe the details of the band structure except for very high-frequency fields or for very small amplitudes,
it gives us good estimates of the parameters needed for switching the band inversion.
Examples of band structures in the first Floquet-Brillouin zone (quasi-energy $\epsilon \in [-\hbar \Omega/2, \hbar \Omega/2]$) are shown in Fig. \ref{fig:bands} calculated with the full Floquet formulation and with $k_x,k_y=0$. In the left panel, the original value of $m_0$ is negative and there are two Dirac points. As we increase the amplitude of the external field with right circular polarization $\eta=1$ and $\hbar \Omega=0.5$ eV, the Dirac points separate and even double, for $\mathcal{A}=0.1,0.2$ \AA$^{-1}$ there are four Dirac points. For $\mathcal{A}=0.3$ \AA$^{-1}$ the four Dirac points have disappeared and gaps have opened in symmetric positions around the origin. For $\mathcal{A}=0.5$ \AA$^{-1}$ the two Dirac points reappear. In the right panel we show the case with positive value of $m_0$ and how the Dirac points are switched on and off again as we increase $\mathcal{A}$. If we would have $\eta=-1$, the Dirac points would get closer to each other and to the origin as we increase $\mathcal{A}$, until they cancel each
  other at the origin and a gap opens. Note that the same band structure, including the Dirac nodes, is repeated periodically as a function of the quasi-energy with period $\hbar \Omega$. Also further nodes appear at higher values of $|k_z|$ (not shown in the figure) corresponding to band crossings with Floquet indices $|n| > 1$.

\vspace{0.5cm}

\begin{center}
{\bf Results for a finite slab}
\end{center}

\begin{figure}
 \includegraphics[width=0.42\textwidth]{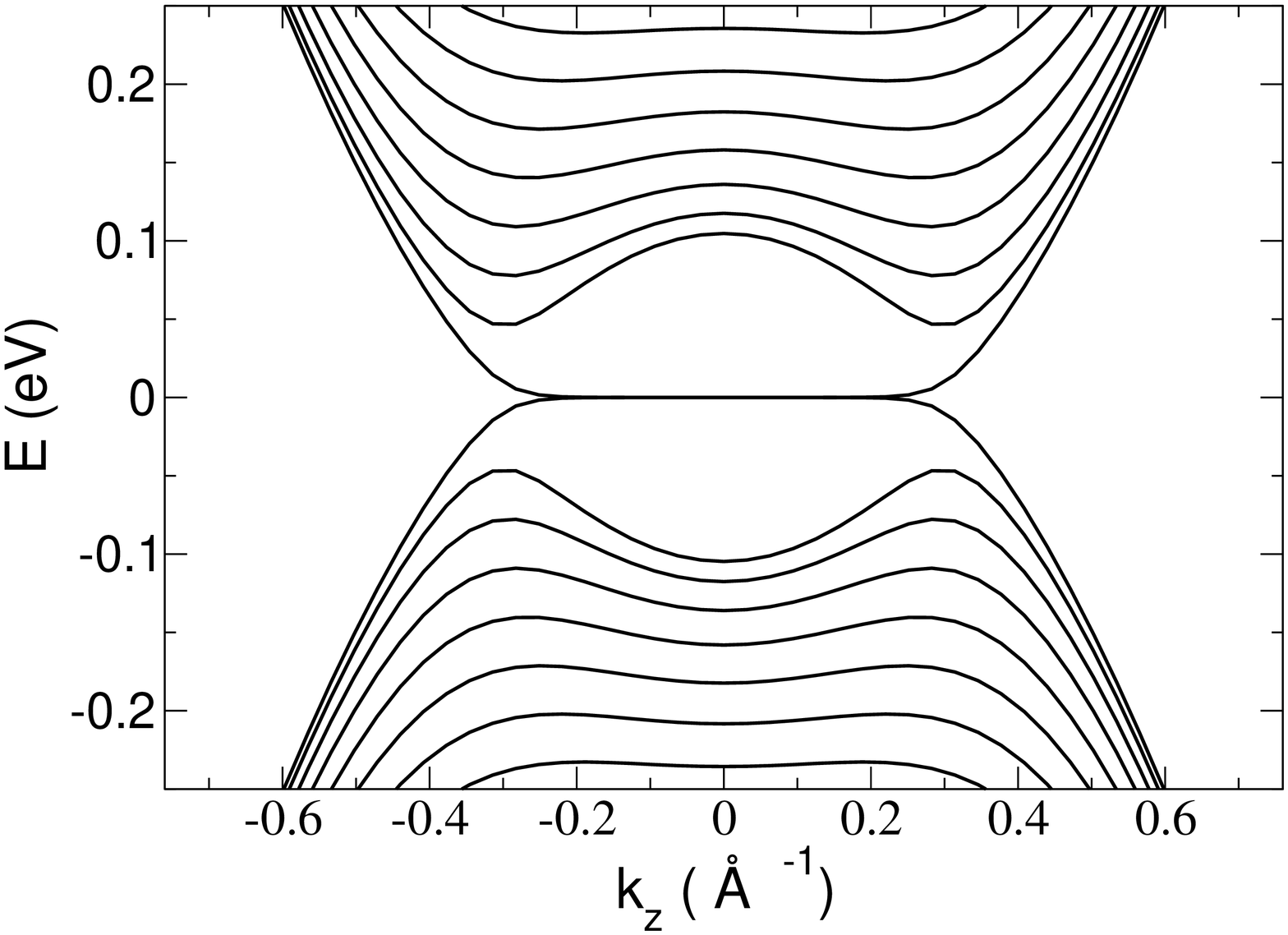}
 \includegraphics[width=0.42\textwidth]{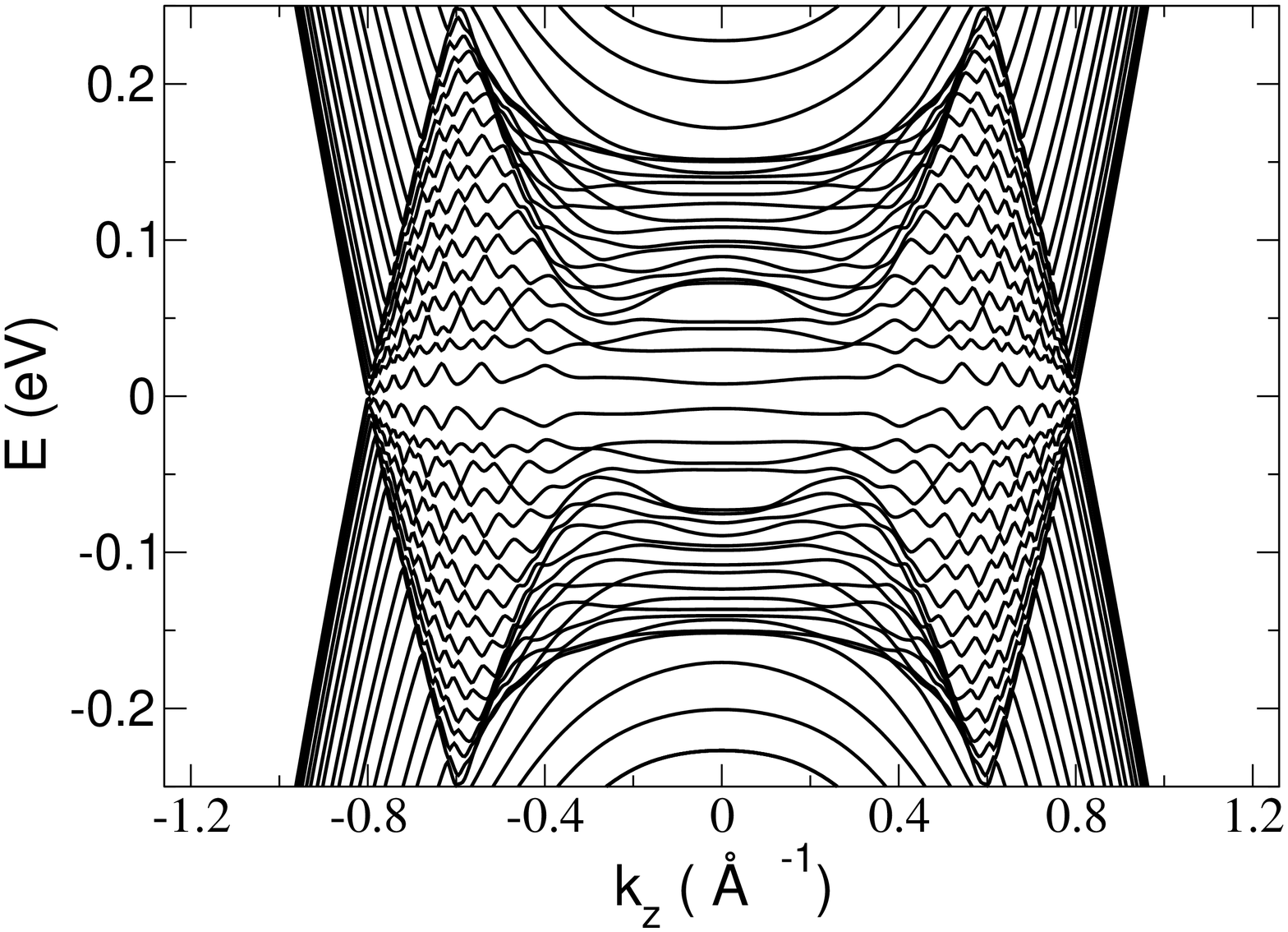}
 \includegraphics[width=0.42\textwidth]{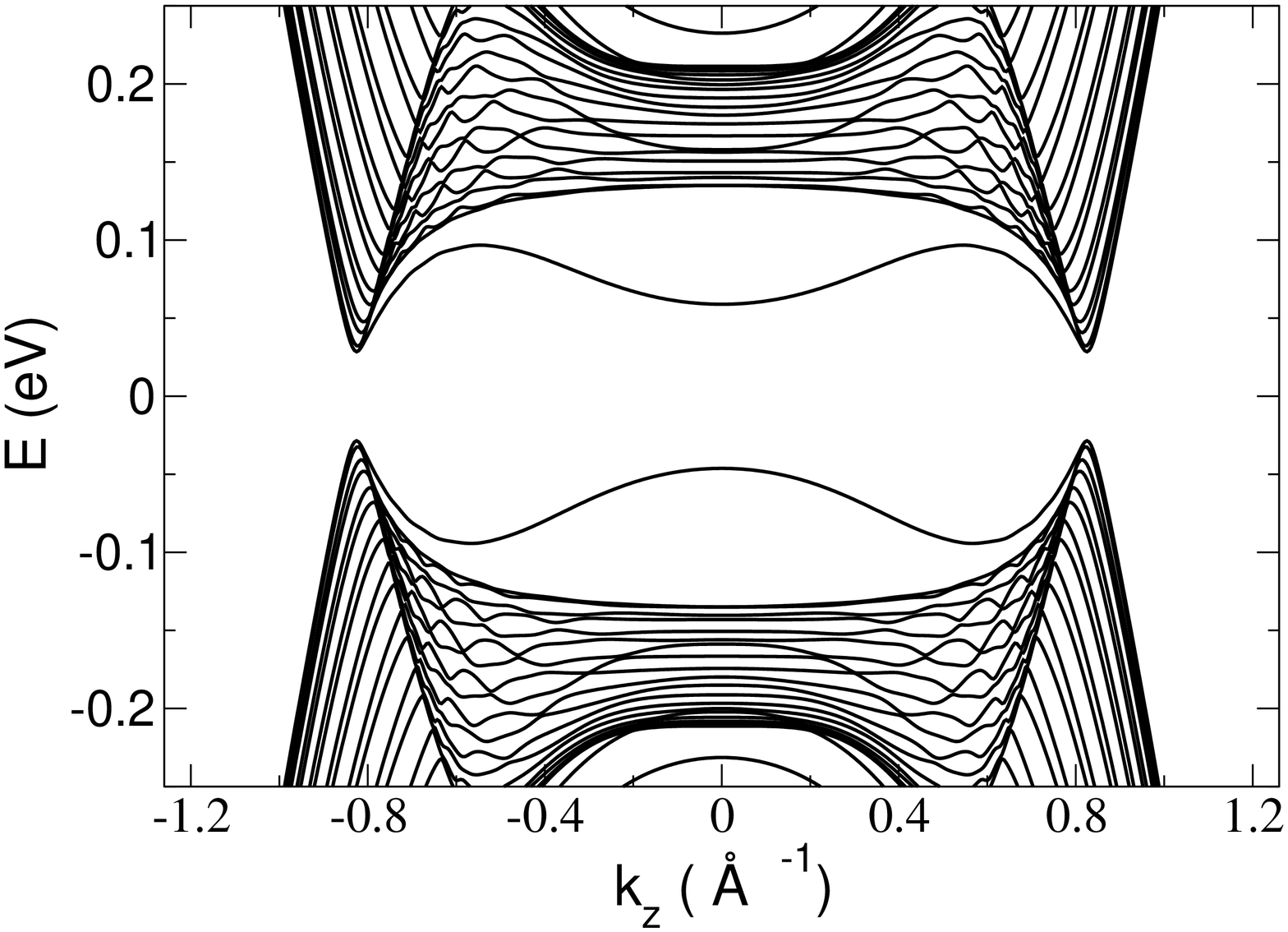}
 \includegraphics[width=0.42\textwidth]{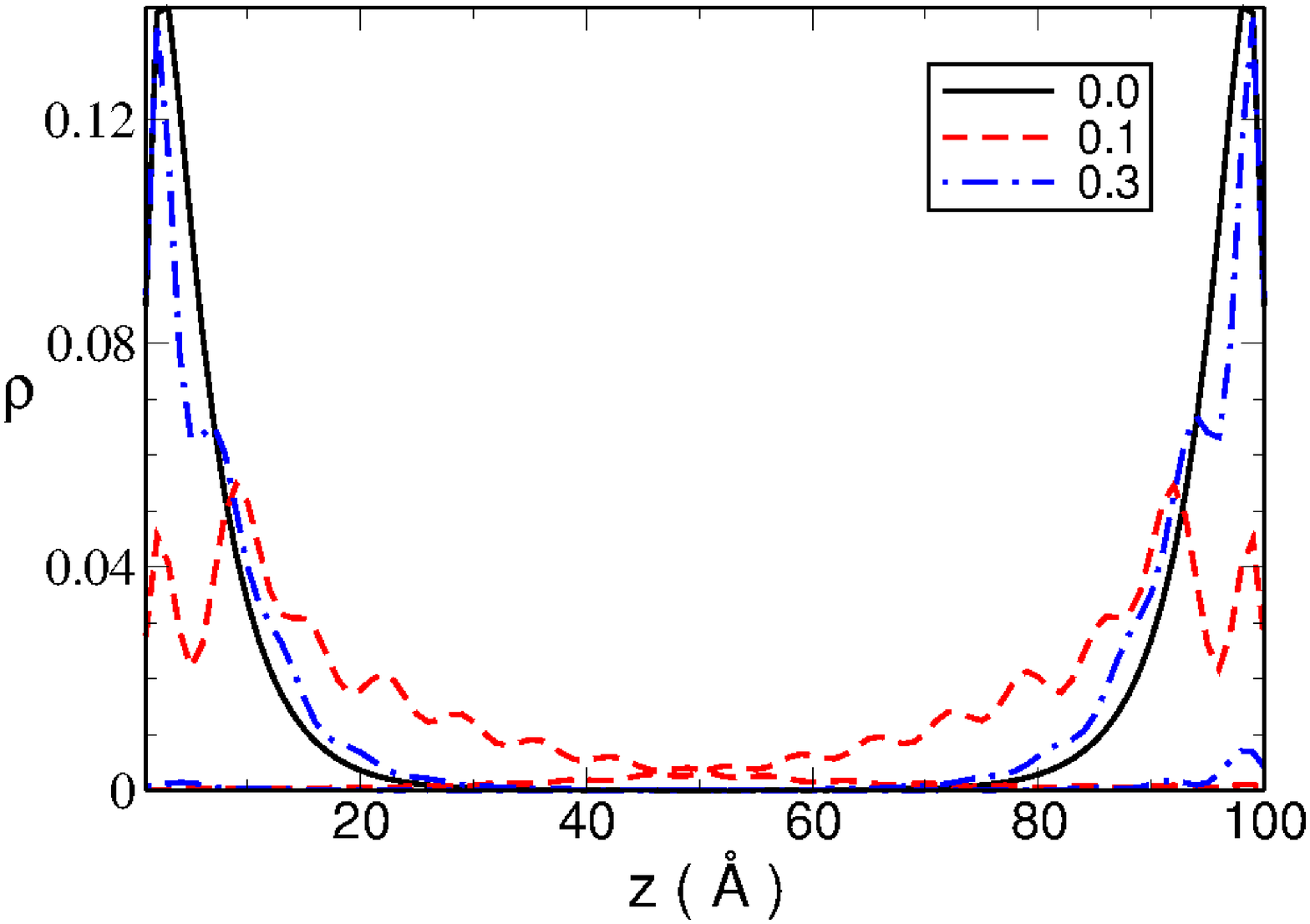}
 \caption{\label{fig:slab} Band structure for a slab of width $W=100$ \AA $ $ in the $x$ direction and infinite in both the $y$ and the $z$ direction as a function of $k_z$ and with $k_y=0$. The Hamiltonian parameters are $c_0=c_1=c_2=0.0$, $m_0=-0.1$ eV , $m_1=m_2=1.0$ eV \AA$^2$, $\hbar v=1.0$ eV \AA, $\zeta=1$, and the external field parameters $\eta=1$, $\hbar \Omega=0.5$ eV, and $\mathcal{A}=0.0$ (top left), 
$0.1$ \AA$^{-1}$ (top right), $0.3$ \AA$^{-1}$ (bottom left). In the right bottom plot, total electron density for states with $k_z=0$ for the different values of $\mathcal{A}$ showing the exponential decay from the surface.}
\end{figure}

To explore the evolution of the midgap Fermi arc states as the external field is switched on, we perform
calculations in a slab of material with finite width $W$ in the $x$ direction but infinite in the $y$ and $z$ directions. We discretize the model Hamiltonian with a standard tight-binding regularization including the Floquet subbands. To make the analysis of the results easier, we present calculations including only one-photon processes, an approximation valid for small values of $\mathcal{A}$ and small values of $k_z$. A more complete analysis of the Full Floquet results is outside the scope of this work and will be presented elsewhere. In Fig. \ref{fig:slab} we show different results of the band structure for a slab of material with the same parameters as the left panel of Fig. \ref{fig:bands} with $\mathcal{A}=0.0,0.1,0.3$ \AA$^{-1}$. We see a band structure resembling the bulk bands calculated previously with midgap states in between the Dirac points. When there is no field, the midgap states form the Fermi arc typical of Weyl semimetals. States in the Fermi arc are surface states as shown in the right bottom panel of the 
figure. Without external field, there are two degenerate Fermi arc surface states, one in each surface. When the field is switched on but we still keep the Dirac points, the Fermi arc states loose the degeneracy between the different surfaces (case with $\mathcal{A}=0.1$ \AA$^{-1}$). For higher fields, when a gap is opened in the Dirac cones, the surface states are not topologically protected anymore (bottom left panel with $\mathcal{A}=0.3$ \AA$^{-1}$). In the bottom left panel we show the total electronic density as a function of the $x$ coordinate for the states with $k_z=0$ for the three values of $\mathcal{A}$ shown previously. There is an exponential decay of the surface states in the Fermi arcs when there is no field and similarly for small values of the field $\mathcal{A}=0.1$ \AA$^{-1}$. For larger fields 
$\mathcal{A}=0.3$ \AA$^{-1}$, the spatial decay from the surface is very much reduced as the gap is opened in the Dirac cones and the topological protection lost.     

\vspace{0.5cm}

\begin{center}
{\bf Signatures of the exceptional points in the complex plane $k_z$}
\end{center}

As mentioned in the main text, the exceptional points in the spectrum of the
Hamiltonian $\widetilde{H}_W $ come in complex conjugate pairs 
$k_i, \overline{k}_i$ in the complex plane of the momentum $k_z$. It may be 
actually seen that the lowest eigenvalue $\varepsilon $ has
along the line connecting each of these pairs the complex structure
\begin{equation}
\varepsilon (k_z) \sim \sqrt{k_i - k_z} \sqrt{\overline{k}_i - k_z}
\label{sr}
\end{equation}

The square root behavior means that $k_i$ and $\overline{k}_i$ are 
branch points and that, for each pair, there are two different branches of 
the Riemann sheet going 
across the real axis of the complex plane. If we take for instance the line 
${\rm Re} (k_z) = {\rm Re} (k_i)$, we get from (\ref{sr})
\begin{eqnarray}
\varepsilon ({\rm Re} (k_i) + iy) & \sim &  \sqrt{i {\rm Im} (k_i) - iy} 
                            \sqrt{-i {\rm Im} (k_i) - iy}      \nonumber   \\
          &  =  &    \sqrt{ ({\rm Im} (k_i))^2 - y^2}
\label{sr2}
\end{eqnarray}
This implies that there must be a segment of points connecting 
$k_i$ and $\overline{k}_i$ (corresponding to $|y| < |{\rm Im} (k_i)|$)
where the eigenvalue $\varepsilon $ becomes purely
real, with two possible opposite values from the two branches in 
the complex plane. On the other hand, when $|{\rm Im} (k_z)| > |{\rm Im} (k_i)|$, 
the above expression also implies the existence of a tail starting from $k_i$ 
(and another from $\overline{k}_i$) where the real part of $\varepsilon $ 
vanishes, while the imaginary part gets opposite values in the two different 
branches.

\begin{figure}[h]
\begin{center}
\includegraphics[height=5.5cm]{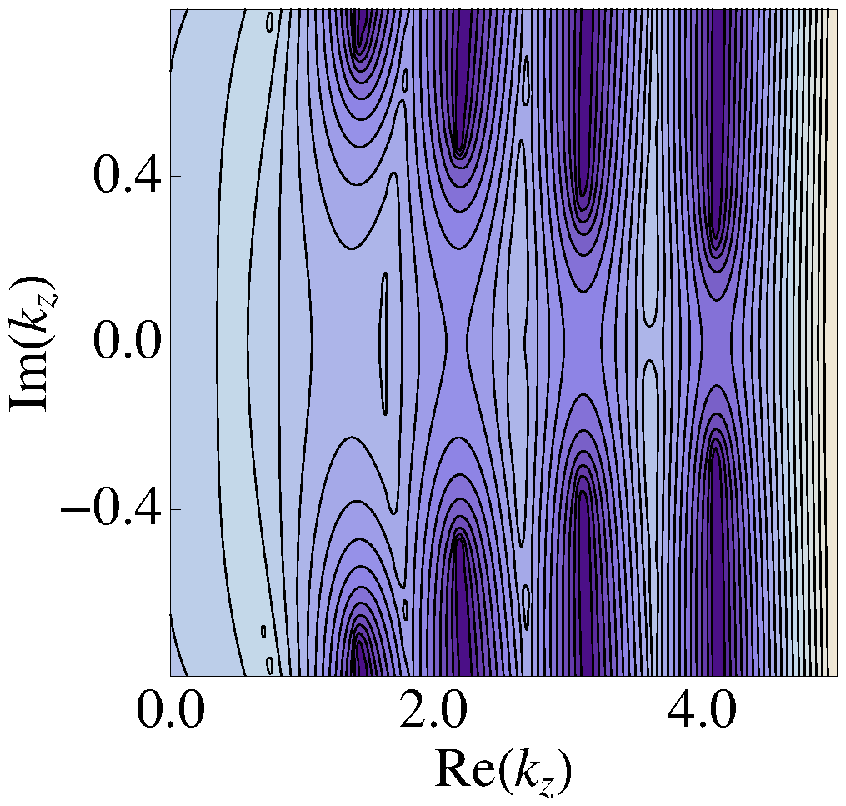}
\hspace{1cm}
\includegraphics[height=5.5cm]{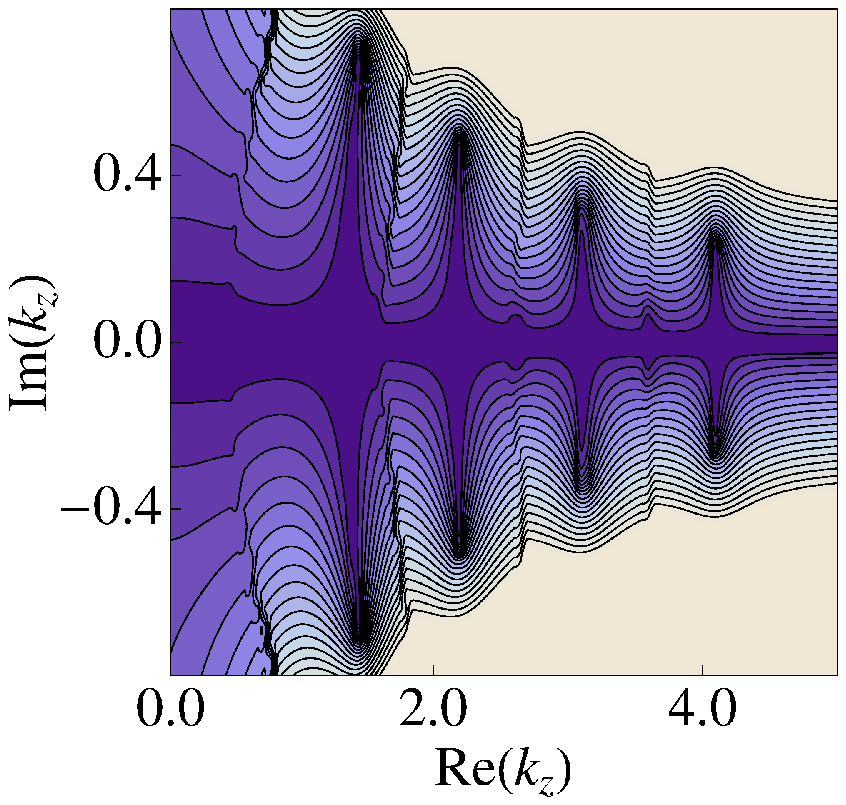}
\end{center}
\caption{Contour plots of the absolute values of the real part (left) and the 
imaginary part (right) of the lowest eigenvalue $\varepsilon $ as a function 
of the momentum $k_z$ in the complex plane, from values close to zero drawn 
in dark blue and larger absolute values shown in light blue. The parameters 
taken for the radiation are 
$\hbar \Omega = 0.1$ eV, $\mathcal{A} = 0.05 \: a^{-1}$, and the radius 
of the cylindrical geometry considered in the main text is $R = 100 a$  
(where $a$ stands for a typical microscopic length scale in the material).}
\label{reim}
\end{figure}

These features can be observed in the results obtained for the lowest 
eigenvalue $\varepsilon $ after numerical diagonalization
of the Hamiltonian, showing the existence of the branch cuts in the spectrum. 
We have represented for instance in Fig. \ref{reim} the absolute values of the 
real and imaginary parts of the eigenvalue $\varepsilon $ for 
radiation parameters $\hbar \Omega = 0.1$ eV, $\mathcal{A} = 0.05 \: a^{-1}$, 
and radius $R = 100 a$ ($a$ standing for a microscopic length scale in the 
material). One can clearly observe the signatures of four pairs 
of exceptional points for ${\rm Re} (k_z) > 0$, which follow to a good 
approximation the behavior expected from (\ref{sr}).

As pointed out in the main text, the number of exceptional points grows 
linearly with the amplitude $\mathcal{A}$ of the vector potential as well as 
with the area of the surface exposed to the radiation. It can be seen in any
case that each pair of complex conjugate exceptional points is tied to the 
same kind of complex structure, which guarantees their robustness as no small 
perturbation can unfold the topology of the underlying Riemann surface.

\vspace{0.5cm}

\begin{center}
{\bf Evaluation of the angular component of the current and 
time evolution of the probability density}
\end{center}

We review here the way in which the probability current ${\bf j}$ of the 
evanescent states can be computed from the spinor 
representation $\chi $ defined in the main text. The formal expression of the 
current ${\bf j}$ can be found multiplying first the evolution equation by the 
complex conjugate of the original Weyl spinor $\psi $, which leads to 
\begin{equation}
i \psi^\dagger \partial_t \psi  =  \psi^\dagger
 [-i v \sigma_x \partial_x -i v \sigma_y \partial_y -i v_z \sigma_z \partial_z 
  + v \mathcal{A} \left(\sigma_x  \cos (\Omega t) + \sigma_y  \sin (\Omega t) \right)]  \psi
\label{ev}
\end{equation}
We can take then the complex conjugate of (\ref{ev}), summing the pair of 
expressions thus obtained to end up with the continuity equation
\begin{equation}
\partial_t \left(  \psi^\dagger \psi \right)  +  
  v \partial_x \left(  \psi^\dagger \sigma_x \psi \right) +
    v \partial_y \left(  \psi^\dagger \sigma_y \psi \right) + 
       v_z \partial_z \left(  \psi^\dagger \sigma_z \psi \right) = 0
\end{equation}
From this expression, we can read the current vector in cartesian 
coordinates
\begin{equation}
{\bf j} =  (v \psi^\dagger \sigma_x \psi , v \psi^\dagger \sigma_y \psi ,
  v_z \psi^\dagger \sigma_z \psi )
\label{cur}
\end{equation}

We are interested in particular in the angular component of the current 
(\ref{cur}), since we want to compute it for the evanescent states which are 
confined to circular rings on the surface exposed to the radiation. 
Passing to polar coordinates $(r,\theta )$, this angular component turns out 
to be
\begin{equation}
j^\theta = - \frac{\sin (\theta)}{r} v \psi^\dagger \sigma_x \psi 
        + \frac{\cos (\theta)}{r} v \psi^\dagger \sigma_y \psi
\label{jang}
\end{equation}
Our goal is to compute $j^\theta $ from the expression of the eigenstates 
obtained in the spinor representation $\chi $.
To perform this operation, we must bear in mind that the spinors $\chi $ are
related to the original Weyl fermions $\psi $ by the consecutive action of
the unitary transformations $U$ and $P$ defined in the main text, so that 
\begin{equation}
\psi = e^{-i \left(-i \partial_\theta + \tfrac{1}{2}\sigma_z \right) \Omega t } \;
         e^{-i \mathcal{A} r \cos (\theta )} \;  \chi
\label{trans}
\end{equation}

For the sake of facilitating the analytic evaluation of the current, we may 
assume at this point the condition $\mathcal{A} r \ll 1 $. We get then 
\begin{eqnarray}
j^\theta  & = & - \frac{\sin (\theta)}{r} 
   v \left( e^{- \Omega t \partial_\theta } \chi^\dagger \right) 
  e^{i \tfrac{1}{2}\sigma_z \Omega t } \sigma_x e^{-i \tfrac{1}{2}\sigma_z \Omega t } 
          \left( e^{- \Omega t \partial_\theta } \chi \right)     \nonumber    \\
      & &     + \frac{\cos (\theta)}{r} 
   v \left( e^{- \Omega t \partial_\theta } \chi^\dagger \right) 
  e^{i \tfrac{1}{2}\sigma_z \Omega t } \sigma_y e^{-i \tfrac{1}{2}\sigma_z \Omega t } 
          \left( e^{- \Omega t \partial_\theta } \chi \right)     \nonumber          \\   
    & = &  - \frac{\sin (\theta)}{r} 
   v \left( e^{- \Omega t \partial_\theta } \chi^\dagger \right)  
    \left(  \sigma_x  \cos (\Omega t)  -  \sigma_y  \sin (\Omega t)  \right)
                \left( e^{- \Omega t \partial_\theta } \chi \right)    \nonumber   \\
    & &    + \frac{\cos (\theta)}{r} 
   v \left( e^{- \Omega t \partial_\theta } \chi^\dagger \right)  
    \left(  \sigma_x  \sin (\Omega t)  +  \sigma_y  \cos (\Omega t)  \right)
                 \left( e^{- \Omega t \partial_\theta } \chi \right)    \nonumber    \\
   & = &  -i \frac{v}{r} \left(  e^{-i(\theta - \Omega t)}
           \left( e^{- \Omega t \partial_\theta } \chi_1^* \right)
              \left( e^{- \Omega t \partial_\theta } \chi_2 \right)
          -    e^{i(\theta - \Omega t)}
           \left( e^{- \Omega t \partial_\theta } \chi_2^* \right)
              \left( e^{- \Omega t \partial_\theta } \chi_1 \right)   \right)
\label{jth}
\end{eqnarray}

We may now work out the expression of the current for a spinor $\chi $
representing the hybridization between states with projection of the total 
angular momentum $j_z = \hbar/2$ and $-\hbar/2$ (omitting at this point the 
common $k_z$-dependence for simplicity)
\begin{equation}
\chi \sim \left( \begin{array}{cc} \phi_1 (r) 
              \\ e^{i\theta} \phi_2 (r) \end{array} \right)  + 
  \left( \begin{array}{cc} e^{-i\theta} \phi_3 (r) 
             \\  \phi_4 (r) \end{array} \right) 
\label{spinor}
\end{equation}
It can be seen that, if we compute $j^\theta $ for each spinor 
contribution with well-defined $j$, the periodic 
time dependence is canceled out in the expression (\ref{jth}).
If instead we carry out the full calculation allowing for the mixing 
of the two spinor contributions, we get a piece in $j^\theta $ that keeps the 
periodic time dependence as in Eq. (\ref{jth}). We obtain the result
\begin{equation}
j^\theta  =  j^\theta_{\rm static}  +  j^\theta_\Omega
\end{equation}
with
\begin{equation}
j^\theta_{\rm static} = -i \frac{v}{r} (\phi_1^* \phi_2 - \phi_2^* \phi_1 
   + \phi_3^* \phi_4  - \phi_4^* \phi_3   )
\end{equation}
and
\begin{equation}
j_\Omega^\theta = -i \frac{v}{r} (e^{-i(\theta - \Omega t)} \phi_1^* \phi_4  
              -  e^{i(\theta - \Omega t)} \phi_4^* \phi_1
                + e^{i(\theta - \Omega t) } \phi_3^* \phi_2  
           -  e^{-i(\theta - \Omega t) } \phi_2^* \phi_3  )
\label{jom}
\end{equation}

\begin{figure}[h]
\begin{center}
\includegraphics[height=3.15cm]{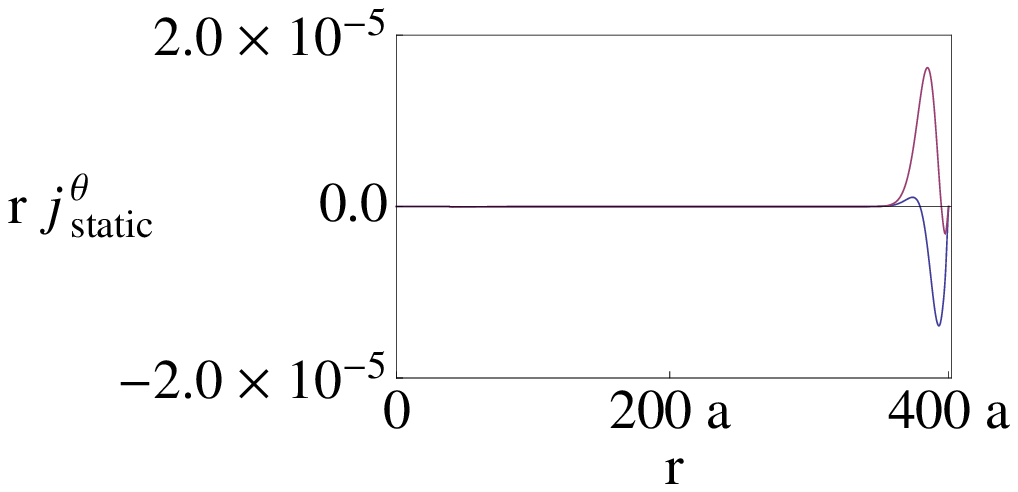}
\hspace{1cm}
\includegraphics[height=3.0cm]{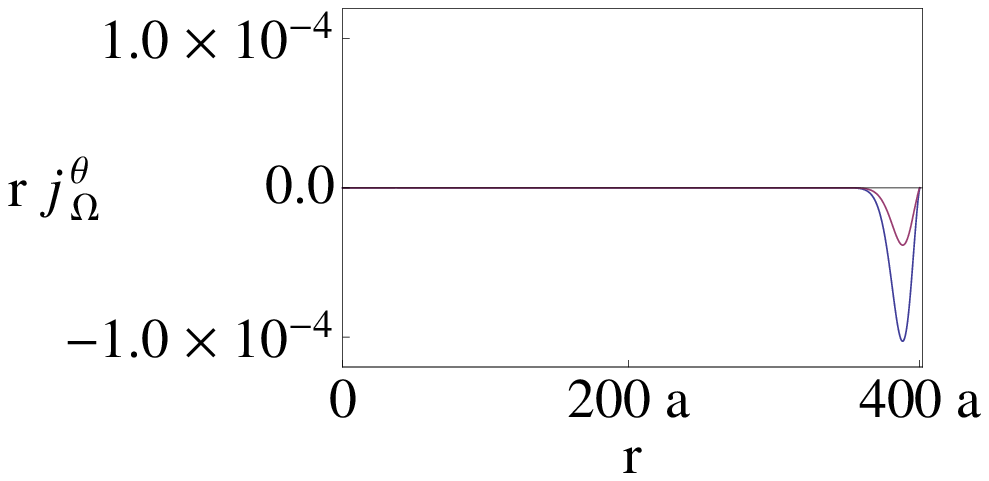}
\end{center}
\caption{Left: Plot of the two contributions to $r j^\theta_{\rm static}$ (in 
units of $v/a^2$, where $a$ stands for a microscopic length scale in the 
material) from the spinors with $j_z = \hbar/2$ (blue) and $j_z = -\hbar/2$ (red) 
making up the evanescent state represented at the top of the left panel in 
Fig. 4 of the main text. 
Right: Plot of the two contributions to $r j_\Omega^\theta $ at $t = 0$, 
$\theta = 0$ (in units of $v/a^2$) from the first two terms (blue) and the last 
two terms (red) in Eq. (\ref{jom}), for the same state as in the left plot. 
The parameters of the radiation are $\hbar \Omega = 0.5$ eV, 
$\mathcal{A} = 0.005 \: a^{-1}$, and the corresponding radius of the 
cylindrical geometry is $R = 400a$.}
\label{jj}
\end{figure}

The static component $j^\theta_{\rm static}$ is made of the contributions from
the two spinors with $j_z = \hbar/2$ and $j_z = -\hbar/2$, which tend to cancel 
each other when integrating the current to obtain the intensity across the 
radial direction. The two different contributions are represented for instance 
in Fig. \ref{jj} for a typical evanescent state at the outermost region of the 
surface exposed to the radiation. The cancellation operating in the integral 
$\int dr \: r j_{\rm static}^\theta $ leads in general to rather small values 
of the intensity associated to $j^\theta_{\rm static}$. This does not apply 
however to the time-dependent component $j_\Omega^\theta$, as shown in Fig.
\ref{jj}. The intensity obtained from this current turns out to have maxima (in 
the angular variable) that are about two orders of magnitude larger than that 
from $j^\theta_{\rm static}$, providing then the main observable signal of the 
evanescent states.

We can express the time-dependent component as  
\begin{equation}
j_\Omega^\theta (r, \theta ) = \frac{v}{r} f(r, \theta - \Omega t)
\end{equation}
in terms of a function with the periodicity 
$f(r, \theta + 2\pi ) = f(r, \theta )$. The continuity equation can be 
approximated then by 
\begin{eqnarray}
\partial_t \left(  \psi^\dagger \psi \right)  & \approx &   
                    -  \partial_\theta j_\Omega^\theta        \nonumber      \\
     &  =  &   - \frac{v}{r} \partial_\theta f(r, \theta - \Omega t )
\label{cont}
\end{eqnarray}
Eq. (\ref{cont}) can be integrated at once to give
\begin{equation}
\psi^\dagger \psi  \approx    \frac{v}{\Omega r}  f(r, \theta - \Omega t )
  +  {\rm const. }
\label{dens}
\end{equation}
This shows that the probability density of the evanescent states has a time
evolution that is inherited from the time dependence of the current.

Similar results are obtained when considering the evanescent states from 
the hybridization of bands with values of the projection of the total angular 
momentum differing by two units of $\hbar $. 
This case is relevant since it corresponds to the appearance of evanescent
states with quasi-energy $\epsilon = 0$ in the conventional Floquet approach.
We may start for instance with a linear combination of states with 
$j_z = \hbar/2, -\hbar/2$ and $-3\hbar/2$
\begin{equation}
\chi \sim \left( \begin{array}{cc} \phi_1 (r) 
              \\ e^{i\theta} \phi_2 (r) \end{array} \right)  + 
        \left( \begin{array}{cc} e^{-i\theta} \phi_3 (r) 
             \\  \phi_4 (r) \end{array} \right)      + 
        \left( \begin{array}{cc} e^{-2i\theta} \phi_5 (r) 
             \\  e^{-i\theta} \phi_6 (r) \end{array} \right)
\label{spin2}
\end{equation}
Inserting (\ref{spin2}) into (\ref{jth}), we observe that the angular current 
can be decomposed now into three different contributions
\begin{equation}
j^\theta  =  j^\theta_{\rm static}  +  j^\theta_\Omega  +  j^\theta_{2\Omega }
\end{equation}
with 
\begin{eqnarray}
j^\theta_{\rm static} & = &  -i \frac{v}{r} (\phi_1^* \phi_2 + \phi_3^* \phi_4  
     + \phi_5^* \phi_6   )   +    {\rm  h. c. }                             \\
j_\Omega^\theta & = &  -i \frac{v}{r} (e^{-i(\theta - \Omega t)} \phi_1^* \phi_4  
              +  e^{i(\theta - \Omega t)} \phi_3^* \phi_2
                + e^{-i(\theta - \Omega t) } \phi_3^* \phi_6  
        +  e^{i(\theta - \Omega t) } \phi_5^* \phi_4  ) + {\rm  h. c. }      \\
j_{2\Omega}^\theta & = &  -i \frac{v}{r} (e^{-2i(\theta - \Omega t)} \phi_1^* \phi_6  
           +  e^{2i(\theta - \Omega t)} \phi_5^* \phi_2  ) +  {\rm  h. c. }
\end{eqnarray}

As shown with the Floquet approach in the main text, the hybridization of the 
bands differing by two units of $\hbar $ in the projection of the total angular 
momentum (energy difference equal to $2\hbar \Omega $) 
opens a gap along a circle in momentum space. It can be seen that the resulting 
evanescent states are confined in concentric circular rings on the surface 
exposed to the radiation, with currents $j^\theta_{\rm static}$ and 
$j_\Omega^\theta $ which have very small intensities 
when integrated along the radial direction. However, the component 
$j_{2\Omega}^\theta $ leads to maxima of the intensity (in the angular 
variable) which are about two orders of magnitude larger than those from 
$j^\theta_{\rm static}$ and $j_\Omega^\theta $, 
producing a significant modulation of the charge which rotates with frequency 
$\Omega $. In this case we have
\begin{equation}
j_{2\Omega}^\theta (r, \theta ) = \frac{v}{r} g(r, \theta - \Omega t)
\end{equation}
in terms of a periodic function with $g(r, \theta + \pi ) = g(r, \theta )$.
The probability density of the evanescent states has now a time evolution 
given by
\begin{equation}
\psi^\dagger \psi  \approx    \frac{v}{\Omega r}  g(r, \theta - \Omega t )
  +  {\rm const. }
\label{dens2}
\end{equation}

From (\ref{dens}) and (\ref{dens2}), we actually conclude that the charge of 
the evanescent states must have an angular modulation and that, moreover, 
it has to rotate with the same frequency $\Omega $ of the radiation, 
along the rings where the states are confined.


\begin{thebibliography}{xx}
\bibitem{vz} 
We set in what follows $v_z = 0.05 v$, taking as a guide the relative magnitudes observed in real 3D semimetals.
\bibitem{f2}
We note that, in the semi-infinite geometry, two modes with momenta $k_z$ and $-k_z$ are to be superposed, in order to build each eigenstate vanishing at the top boundary of the cylinder.
\bibitem{f1}
A typical magnitude of the imaginary part of the EPs is for instance ${\rm Im} (k_i) \sim 0.1 \: a^{-1}$, for $\hbar \Omega = 0.5$ eV and $\mathcal{A} = 0.005 \: a^{-1}$, where $a$ stands for a typical microscopic length scale 
in the material.
\bibitem{f3}
Our definition of the current in Eq. (\ref{jomega}) (as well as that of $j_{\rm static}^\theta $) omits for simplicity its decay along the $z$-axis, which is inherited from the evanescence of the surface states.
\end{thebibliography}
\end{document}